\documentclass[aps,prl,preprint,superscriptaddress]{revtex4-2}
\usepackage{amsmath}
\usepackage{graphicx}
\usepackage[colorlinks=true, allcolors=blue]{hyperref}
\usepackage{lineno}
\usepackage{multirow}

\begin{document}


\title{A Practical Partial-Wave Method for Identifying Unstable Light Nucleus Resonances in Heavy-Ion Collisions}


\author{Junlin Wu}
\email[]{wujunlin@ucas.ac.cn}
\affiliation{School of nuclear Science and Technology, University of Chinese Academy of Sciences, Beijing, 101408, China}

\author{Hongchan Li}
\affiliation{Key Laboratory of Quark and Lepton Physics (MOE) and Institute of Particle Physics, Central China Normal University, Wuhan, 430079, China}

\author{Ke Mi}
\affiliation{School of nuclear Science and Technology, University of Chinese Academy of Sciences, Beijing, 101408, China}

\author{Yaping Wang}
\affiliation{Key Laboratory of Quark and Lepton Physics (MOE) and Institute of Particle Physics, Central China Normal University, Wuhan, 430079, China}

\author{Guannan Xie}
\email[]{xieguannan@ucas.ac.cn}
\affiliation{School of nuclear Science and Technology, University of Chinese Academy of Sciences, Beijing, 101408, China}


\date{\today}

\begin{abstract}

The production of light nuclei in relativistic heavy-ion collisions provides valuable insights into the dynamics of the hot and dense matter created in these extreme environments. While stable light nuclei have been extensively studied, unstable light nuclei being short-lived resonance states remain largely unexplored and offer unique opportunities to probe final-state interactions and the freeze-out conditions. In this paper, we propose a partial-wave method, based on the Lednick\'y--Lyuboshitz framework, to extract resonance signals of unstable light nuclei from two-particle correlation functions measured in heavy-ion collisions. By extending the LL model to higher order partial waves and directly incorporating experimental phase-shift data from low-energy nuclear scattering, our approach avoids the need for model-dependent potential parametrizations and enables a clean decomposition of the resonant partial wave from the non-resonant background. As a demonstration, we apply the method to the $p$-$^3$He and $p$-$^4$He systems, corresponding to the $^4$Li and $^5$Li ground-state resonances. Numerical results show that the resonance-induced correlation excess can be effectively isolated, with a peak in the correlation function appearing at $k \approx 72$ MeV/$c$ for $^4$Li and $k \approx 50$ MeV/$c$ for $^5$Li, consistent with the known resonance parameters. The extracted transverse momentum spectra and rapidity distributions are presented using the measured proton and light-nuclei spectra from STAR at $\sqrt{s_{NN}} = 3$ GeV. The proposed method provides a practical tool for the experimental study of unstable light nuclei in relativistic heavy-ion collisions and can be extended to a broader range of resonance states.
\end{abstract}


\maketitle

\section{Introduction}

The study of light nuclei production in relativistic heavy-ion collisions (HICs) has emerged as a crucial tool for probing the dynamics of the hot and dense matter created in these extreme environments. Light nuclei, being loosely bound clusters of nucleons, are sensitive to the conditions at the kinetic freeze-out stage, providing constraints on the freeze-out temperature, volume, and collective flow. Over the past decades, significant progress has been made in measuring stable light nuclei, such as deuterons ($d$), tritons ($t$), $^3\mathrm{He}$, and $^4\rm{He}$ particles, in experiments at the Relativistic Heavy Ion Collider (RHIC) and the Large Hadron Collider (LHC) \cite{STAR:2019sjh,STAR:2022hbp,STAR:2023uxk,Jin:2025biy,ALICE:2015wav,ALICE:2017xrp,ALICE:2017jmf,ALICE:2020foi,ALICE:2019bnp,ALICE:2021mfm,ALICE:2021ovi,ALICE:2025byl}. 

These measurements have been described by coalescence models \cite{Bazak:2018hgl} and statistical thermal models \cite{Andronic:2010qu}. However, both face certain challenges. The thermal model regards the heavy-ion collision system as a grand canonical ensemble, where the ensemble is determined by the temperature $T$, the baryon chemical potential $\mu_B$, and the volume of the system $V$. All reaction channels reach equilibrium and the yields of the particles are almost determined by their masses and spin degeneracy factors. For the light nuclei produced by the thermal model, there arises a well-known paradox: the "snowball in the furnace". The binding energy of light nuclei is only a few MeV. It is hard to imagine that they could exist in the collision system with a freeze-out temperature of around 100 MeV. The explanation of the thermal model for the paradox is that it is entropy conservation that governs the production yields of light nuclei, but not the difference between the binding energy and temperature of the system. Such an explanation is fully supported by thermodynamics, while it lacks a microscopic mechanism. 

To solve this paradox, the coalescence model suggests that light nuclei are formed in the later stages of the collision system's evolution. When the coordinates and momenta of several nucleons are very close to each other, there is a probability for them to coalesce into a light nucleus. Such calculations are usually performed in conjunction with hadron transport models (JAM, UrQMD, etc.) and are referred to as afterburner coalescence. This method relies on the hadron transport model and a few parameters need to be determined, such as the coalescence time and the coalescence probability distribution function. However, it exhibits a high sensitivity to the dynamic parameters. Moreover, even the simplest coalescence process $p + n \rightarrow d$ is also an equivalent physical process, otherwise it violates momentum conservation. Therefore, we believe that the coalescence process is merely an effective description. It should effectively represent all low-energy nuclear reactions that produce this type of light nucleus, which occur via final-state interactions in heavy-ion collisions.

Studying unstable light nuclei can help us understand this issue. Taking $^{4}\rm{Li}$ (g.s.) and $^{5}\rm{Li}$ (g.s.) as examples, the mass of $^{4}\rm{Li}$ exceeds the $p+{}^3\mathrm{He}$ mass by only 4.07 MeV, with a width of approximately 5.5 MeV, corresponding to a lifetime of about 30 fm/$c$, and $J^{\pi}=2^{-}$. The mass of $^{5}\rm{Li}$ is only higher than the mass of $p+ {}^{4}\rm{He}$ by 1.69 MeV, with a width of approximately 1.5 MeV, corresponding to a lifetime of about 100 fm/$c$, with $J^{\pi}=\frac{3}{2}^{-}$. Both decay by proton emission, which means that they are the resonant states of the $p{}^{3}\rm{He}$ or $p{}^{4}\rm{He}$ system. Among all the low-energy nuclear reactions that could produce $^{4}\rm{Li}$ and $^{5}\rm{Li}$, except for elastic scattering($p+ {}^{3}\rm{He}\rightarrow p+ {}^{3}\rm{He}$ and $p+ {}^{4}\rm{He}\rightarrow p+ {}^{4}\rm{He}$), either the reactants require a higher nucleon number $A$ or the reaction requires to absorb energy (such as ${}^{3}\rm{He}+{}^{3}\rm{He}\rightarrow d +p+ {}^{3}\rm{He}$, ${}^{4}\rm{He}+d\rightarrow n +p+ {}^{4}\rm{He}$, ${}^{6}\rm{Li}+\gamma\rightarrow n +p+ {}^{4}\rm{He}$...). In the heavy-ion collision environment, these channels are expected to be suppressed compared to elastic scattering\cite{Dong:2020hxe,Zhang:2024qkg}.

There are other motivations for studying unstable light nuclei. In non-central relativistic heavy-ion collisions, the spin-orbit coupling may induce polarization or spin alignment of the emitted particles. RHIC-STAR have measured the $\Lambda$($\bar{\Lambda}$) global polarization\cite{STAR:2017ckg} as well as $\phi$ meson spin alignment\cite{STAR:2022fan} by measuring the angular distribution of the final states of the daughter particles. A natural question is how protons and light nuclei are polarized.The spin alignment of $^{4}\rm{Li}$ could provide insight into this question\cite{Zheng:2025ngn}.

Searching for heavier antimatter in heavy-ion collision experiments helps us understand the origin of the asymmetry between matter and antimatter in the universe. The currently discovered antimatter nuclei in heavy-ion collisions include $^{3}_{\bar{\Lambda}}\overline{\rm{H}}$, $^{4}\overline{\rm{He}}$, $^{4}_{\bar{\Lambda}}\overline{\rm{H}}$, and $^{4}_{\bar{\Lambda}}\overline{\rm{He}}$. The next stable antimatter nucleus is $^6\overline{\rm{Li}}$, which has about seven to eight orders of magnitude lower yield than that of $^{4}\overline{\rm{He}}$ \cite{STAR:2011eej} due to the penalty factor. With current experimental methods, this is almost impossible to achieve. The discovery of $^4\overline{\rm{Li}}$, however, can be realistically expected \cite{Xi:2019vev}.

Despite the strong motivations, extracting the signal of unstable light nuclear resonance states from the invariant mass distribution is challenging.
In experimental measurements, the difficulty lies in the fact that the invariant mass signal spectrum of unstable light nuclei is often mixed with the contributions of other partial waves. Measurements of unstable light nuclei in heavy-ion collisions are extremely rare. Only the E864 Collaboration has carried out pioneering work\cite{Armstrong:2001mr} in this area. Due to the limitations of experimental accuracy, that work did not consider partial wave analysis.

In order to illustrate the necessity of partial wave analysis in the measurement of unstable light nuclei, we need to review the discovery of $^{4}\rm{Li}$. More than half a century ago, $^{4}$Li was discovered in low-energy nuclear scattering experiments, such as $p + {}^{3}\mathrm{He} \rightarrow p + {}^{3}\mathrm{He}$ elastic scattering~\cite{Reichstein:1971wa,Tombrello:1965zz,Alley:1993zza,Alley:1993zz}. By measuring the scattering cross-section and polarization observables, a phase-shift analysis was performed. A resonance manifests itself as a rapid variation of the phase shift as a function of energy. In the $p$-$^{3}$He system, the $^{3}P_2$ partial wave exhibits such a rapid change, corresponding to the peak of the partial wave scattering cross-section, and it is identified as the ground state of $^{4}$Li~\cite{Tombrello:1965zz}. From this perspective, $^{4}$Li is a resonance in the $p$-$^{3}$He system. The mass of $^{4}$Li is only slightly above the $^{3}$He + $p$ threshold, so the $p$-$^{3}$He system is naturally probed in the low relative momentum region. In heavy-ion collisions, particles are emitted from a common source and may interact after freeze-out. These final-state interactions (FSI) occur at low relative momentum and thus correspond to the same low-energy nuclear scattering processes. Importantly, near the threshold region in the invariant mass of $p$-$^{3}\rm{He}$, \textbf{other partial waves}---such as $^{1}S_0$, $^{3}S_1$, $^{1}P_1$, $^{3}P_0$, and $^{3}P_1$---also contribute. These non-resonant partial-wave contributions are the origin of the correlated background in the invariant mass distribution. Our goal is therefore to separate the resonant $^{3}P_2$ (i.e., $^{4}$Li) contribution from the other partial-wave contributions.

The correlation function is defined as a function of the relative momentum $k$ in the two-particle center-of-mass system, representing the ratio of the probability of detecting a particle pair to that in the absence of interactions. The experimental correlation function is defined as
\begin{equation}
    C(k)=\mathcal{N}\frac{N_{pair}(k)}{N_{bkg}(k)},
\end{equation} 
where $N_{pair}(k)$ is the counts distribution of the measured particle pairs, and $N_{bkg}(k)$ is the reference counts distribution of the measured particle pairs without interaction. In experiments, $N_{bkg}(k)$ is usually generated by the mixed event method or the rotation method. $\mathcal{N}$ is a normalization parameter. 
Under a fixed particle mass, invariant mass $M_{\rm{inv}}$ has a one-to-one kinematic mapping with $k$. Thus the invariant mass spectrum and the momentum correlation function represent the same physical information in different ways. The correlation function is better suited for displaying the structure near the threshold, while the invariant mass spectrum is more conducive to extracting the number of signals. In practice, the two approaches can be used complementarily to cross-check the resonance signal. For a resonance, the correlation function exhibits a characteristic enhancement at a relative momentum corresponding to the resonance energy.

STAR at $\sqrt{s_{NN}}=3$ GeV and ALICE at LHC energies have reported the first measurements of proton-deuteron (p-d) and deuteron-deuteron (d-d) correlation functions \cite{STAR:2024zvj,ALICE:2025wuy}. The source size, the scattering length and effective range have been extracted from the Lednick\'y--Lyuboshitz (LL) model \cite{Lednicky:1981su} by fitting the data. The extracted scattering length and effective range were compared with the results of low-energy nuclear scattering experiments. However, the interaction between these two pairs is dominated by the repulsive s-wave and does not involve resonances. When the pairs involve heavier light nuclei, the correlation functions will show the complex structure of resonant states and higher-order partial waves. The traditional LL model, which is primarily formulated for s-wave interactions, is insufficient for extracting resonant signals from these systems. This motivates our extension of the LL model to arbitrary partial waves.

Several previous studies have attempted to extract resonance signals from correlation functions using different approaches. In Ref.~\cite{Xi:2019vev,Bazak:2020wjn}, the $^4$Li signal was manually inserted into the correlation function. The $p$-$^3$He $P$-wave contribution was modeled by a Breit-Wigner amplitude, requiring fitting parameters for the width, peak position, and source radius, but the spin degrees of freedom were not discussed. Recently, Murase and Hyodo \cite{Murase:2025nlo} developed a general formalism for higher partial waves in femtoscopy by introducing a cutoff regularization to cure the divergence of the asymptotic wave function. However, their work is restricted to the case without Coulomb interaction. 
In this paper, we extend their regularization scheme to include the full Coulomb interaction, which is essential for charged systems such as $p$-$^3$He and $p$-${}^{4}\rm{He}$. We use experimental phase shifts as input for the exterior scattering wave rather than constructing a parameterized nuclear potential, and we provide a practical framework for extracting resonance signals from experimental correlation functions. We propose a partial-wave method, based on the LL model \cite{Lednicky:1981su}, to extract the signals of unstable light nuclei from two-particle correlation functions measured in HICs. We extend the LL model to higher order partial waves by directly incorporating phase-shift data from low-energy nuclear scattering experiments. This approach avoids the need for detailed potential parametrizations and allows a decomposition of the resonant partial wave from the background. As a demonstration, we apply our method to the specific cases of $^{4}\mathrm{Li}$ (from the $p$-$^3\mathrm{He}$ system, $^{3}P_2$ resonance) and $^{5}\mathrm{Li}$ (from the $p$-$^4\rm{He}$ system, $P_{3/2}$ resonance). Our numerical results show that the resonant signals can be effectively isolated, providing a practical tool for experimental studies of unstable light nuclei in heavy-ion collisions.

The paper is organized as follows. In Sec.~II, we present the theoretical framework, including the Koonin-Pratt equation and the extension of the LL model to include higher partial waves and Coulomb interactions. In Sec.~III, we describe the numerical implementation and present our results for the $^{4}\mathrm{Li}$ and $^{5}\mathrm{Li}$ correlation functions. A summary and outlook are given in Sec.~IV.

\section{General formula and Lednicky-Lyuboshitz model}
Theoretically, the correlation function can be calculated via the Koonin-Pratt equation\cite{Koonin:1977fh,Lednicky:1981su}
\begin{equation}
     C(k)=\int d^{3}r S(\vec{r}) \vert \psi(\vec{r},\vec{k}) \vert^{2}
    \label{eq:kp}
\end{equation}
where $\vec{k}$ is the relative momentum in the two-particle center-of-mass frame, and $\vec{r}$ is the relative distance between the two particles when they are at the emission source. $S(\vec{r})$ is the emission source function. 
The LL approach \cite{Lednicky:1981su} is used to parameterize the measured two-particle correlation function. The LL approach relies on three assumptions: the equal-time emission approximation, the asymptotic wave function, and the scattering amplitude. In this paper, a spherical Gaussian distribution is adopted:
\begin{equation}
    S(r)= \frac{1}{(4\pi R_g^2)^{\frac{3}{2}}}e^{-\frac{r^{2}}{4R_g^2}}
    \label{eq:gauss_source}
\end{equation}
where $R_g$ is the radius of the emission source function which is determined by fitting the experimental data. Here, the equal-time approximation is employed, that is, the two particles are emitted simultaneously from the source. Although the spherical Gaussian distribution is a relatively strong assumption, it is still a reasonable approximation within the accuracy of most analyses \cite{Lednicky:1981su,STAR:2024zvj,STAR:2025jwe,STAR:2015kha,STAR:2018uho,STAR:2014dcy,STAR:2014dcy}.

Strictly, it is necessary to solve the Schr\"odinger equation to obtain the two-particle wave function $\psi$ in Eq.~\ref{eq:kp}. This requires the nuclear force potential in the Hamiltonian, which is usually parameterized \cite{Viviani:1998gr,Pudliner:1997ck,Wiringa:1994wb,Mihaylov:2018rva,Ge:2025put}, and the phase shift data still need to be fitted to determine the parameters. If the nuclear force potential is reliable, this method should lead to the most accurate results.  

Since the nuclear force is short-range, with a typical range smaller than $\sim 1$ fm, the LL approach advocates the use of the asymptotic wave function. It avoids solving the complex differential equations and has also achieved success in many analyses of two-particle correlation functions \cite{Lednicky:1981su,STAR:2024zvj,STAR:2025jwe,STAR:2015kha,STAR:2018uho,STAR:2014dcy,STAR:2014dcy} in HIC. The asymptotic wave function can be expressed as several parts:
\begin{equation}
    \psi=\psi_{in}+f\cdot\psi_{out}
\end{equation}
$\psi_{in}$ represents the ingoing wave, $\psi_{out}$ represents the outgoing wave after scattering, and $f$ is the scattering amplitude.

The traditional LL approach is only applicable to s-waves \cite{Lednicky:1981su}. In other words, $\psi_{out}$ is the eigenfunction of the angular momentum with $l=0$. However, unstable light-nucleus resonances are usually not s-wave, so we will extend it to any partial wave in the form of 
\begin{equation}
    \psi=\psi_{in}+\sum_{l}f_l\cdot\psi_{out}^{l}
\end{equation}
in this paper.

Next, we will separately discuss the asymptotic wave function and the scattering amplitude without Coulomb interaction and with Coulomb interaction.

\subsection{Asymptotic wave function without Coulomb interaction}
To compute the correlation function from Eq.~\ref{eq:kp}, we follow the LL approach framework, using plane waves as $\psi_{in}$:
\begin{equation}
    \psi_{in}=e^{ikz}
    \label{eq:pl_wave}
\end{equation}
If we disregard any scattering between the two particles caused by any interactions, i.e., $\psi=\psi_{in}$, and substitute Eq.~\ref{eq:pl_wave} into the KP equation Eq.~\ref{eq:kp}, the result is obviously $C(k)=1$. The plane wave Eq.~\ref{eq:pl_wave} can be expanded into the eigen wave functions of angular momentum:
\begin{equation}
    \psi=e^{ikz}=e^{ikr\cos{\theta}}=\sum_{l=0}^{\infty}(2l+1)i^{l}j_{l}(kr)P_{l}(\cos{\theta})
    \label{eq:pl_wave_expand}
\end{equation}
where $P_{l}(\cos{\theta})$ is the Legendre polynomial of order $l$, and $j_{l}(kr)$ is the spherical Bessel function of the first kind. By using the orthogonality property of these functions, it can be seen that the correlation function $C(k^*)=1$ is composed of contributions from various partial wave components, even though there is no interaction between the two particles:
\begin{equation}
    C(k)=\sum_{l=0}^{\infty}4\pi(2l+1)\int d^3r S(r) \vert j_{l}(kr)\vert^{2}=1
\end{equation}  
Here, it includes the integral over the scattering angle $\theta$. An integral identity 
\begin{equation}
    \int d\Omega P_{l}(\cos\theta)P_{l'}(\cos\theta)=\frac{4\pi}{2l+1}\delta_{ll'}
\end{equation}
is employed.
In the case where there is a central force field interaction, the eigen wave function of angular momentum is not $j_{l}(kr)$ but $R_{l}(kr)$. Here $R_l$ is the solution of the radial Schr\"odinger equation
\begin{equation}
    \left[ \frac{1}{r^2}\frac{d}{dr}r^2\frac{d}{dr} +k^2 -\frac{l(l+1)}{r^2}-2m_{red}V_N(r) \right]R_l=0
    \label{eq:rad_sch_eq}
\end{equation}
where $m_{red}=m_1m_2/(m_1+m_2)$ is the reduced mass of the two particles. When $V(r)=0$, $j_l(kr)$ is a possible form of $R_l$. Since the nuclear force potential is usually parameterized and depends on the orbital angular momentum and the spins of the two particles, directly solving Eq.~\ref{eq:rad_sch_eq} is rather complicated. According to the spirit of the LL model and scattering theory, each component $(2l+1)i^{l}j_{l}(kr)P_{l}(\cos{\theta})$ of the plane wave is scattered into a spherical wave of the form $f(k,\theta)\frac{e^{ikr}}{r}$ at infinity. All we need to know is the asymptotic wave function when $r\rightarrow\infty$. Utilizing the asymptotic properties of Bessel functions, we take
\begin{equation}
\begin{aligned}
    \psi&=\sum_{l=0}^{\infty}(2l+1)i^{l}R_{l}(kr)P_{l}(\cos{\theta})\\
    &\xrightarrow[]{r\rightarrow\infty}e^{ikz}+\sum_{l=0}^{\infty}f_l(k,\theta)\frac{e^{ikr}}{r}
    \label{eq:plane_spherical_wave}
\end{aligned}
\end{equation}
where 
\begin{equation}
    R_l(kr)\xrightarrow[]{r\rightarrow\infty}j_l(kr)+\frac{e^{2i\delta_l}-1}{2}h_l(kr)
\end{equation}
$h_l(kr)$ is the Hankel function and $\delta_l$ represents the phase shift of each partial wave. The partial wave scattering amplitude is
\begin{equation}
    f_l(k,\theta)=(2l+1)\frac{1}{k(\cot\delta_l-i)}P_{l}(\cos{\theta})
\end{equation}
Since the phase shifts introduce an additional scattering term, the correlation function deviates from 1:
\begin{equation}
    C(k)=1+\sum_{l=0}^{\infty}\Delta C_l(k)
\end{equation}    

\begin{equation}
    \Delta C_l(k)=4\pi(2l+1)\int d^3r S(r) M_l(kr)
\end{equation}
where
\begin{equation}
    M_l(kr)= \vert j_l(kr)+\frac{e^{2i\delta_l}-1}{2}h_l(kr)\vert^2 - \vert j_l(kr)\vert^2 
\end{equation}
In particular, when $l=0$, a Taylor expansion of even powers for $k\cot\delta_0$ can be used for parameterization:
\begin{equation}
    f_0(k,\theta)=\left(-\frac{1}{a_0}+\frac{r_0}{2}k^2-ik\right)^{-1}
    \label{eq:effR_noC}
\end{equation}
Where $a_0$ is scattering length and $r_0$ is effective range. This is the form of the traditional LL model.

For simplicity, we avoid solving the Schrödinger equation for nuclear force parametrization as much as possible. Rather, within the framework of the LL model, the phase shift data measured in low-energy nuclear scattering experiments are used as the input for the asymptotic wave function. When the size of the emission source is large, the success of the LL model \cite{Lednicky:1981su,STAR:2024zvj,STAR:2025jwe,STAR:2015kha,STAR:2018uho,STAR:2014dcy,STAR:2014dcy} indicates that an accurate wave function is not always necessary. The asymptotic wave function can describe the data most of the time.

\subsection{Asymptotic wave function with Coulomb interaction}
It is difficult to precisely measure the momentum of neutrons in particle detectors, and thus it is challenging to measure the correlation functions between neutrons and other light nuclei. For instance, the observation of the $^5$He resonance state in the $^5$He-n correlation function is almost impossible. Therefore, in the analysis of experimental data, situations involving Coulomb interactions are more common. 
The system potential energy is as follows:
\begin{equation}
    V(r)= V_N(r) + \frac{Z_1 Z_2}{r} e^{2}
\end{equation}
where $V_N(r)$ represents the short-range nuclear force potential, which typically only acts within a range of less than $\sim 1$ fm. $Z_1$ and $Z_2$ represent the charge numbers of the two particles. The Coulomb interaction is of the form $1/r$ and is a long-range force. Being a long-range force, it leads the asymptotic wave function at infinity to include the contributions of all partial waves. 
Fortunately, the Coulomb interaction is analytically solvable. By approximating the wave function beyond the range of the nuclear force, the problem can be solved. In this case, the well-known Coulomb wave function is given by the following formula \cite{Lednicky:1981su}:
\begin{widetext}
\begin{equation}
    \psi(\vec{r},\vec{k})=\sqrt{A_c(\eta)}e^{i\delta^{C}_{0}}[ e^{i\vec{k}\cdot \vec{r}}F(-i\eta,1,ikr-i\vec{k}\cdot\vec{r})+\sum_{l=0}^{\infty}f^{NC}_l(k,\theta) \frac{\tilde{G_{l}}(\rho,\eta)}{r}]
    \label{eq:coulomb_wave}
\end{equation}
\end{widetext}
Where $e^{2i\delta^{C}_{l}}=\Gamma(l+1+i\eta)/\Gamma(l+1-i\eta)$, and $\delta^{C}_{l}$ is the Coulomb phase shift of order $l$. $A_c(\eta)=2\pi\eta/(e^{2\pi\eta}-1)$ is the Coulomb penetration factor, and $\eta=Z_1 Z_2 m_{red} e^2/k$ is the Sommerfeld parameter. $F$ is the confluent hypergeometric function. And $\tilde{G_l}(kr,\eta)$ is the linear combination of the Coulomb function of the first kind $F_l(kr,\eta)$ and the Coulomb function of the second kind $G_l(kr,\eta)$:
\begin{equation}
    \tilde{G_l}(kr,\eta)=\sqrt{A_c(\eta)}(G_l(kr,\eta)+iF_l(kr,\eta))
\end{equation}
The forms of Eq.~\ref{eq:coulomb_wave} and Eq.~\ref{eq:plane_spherical_wave} are similar, except that both the plane wave and the spherical wave have been distorted by the Coulomb field.
The partial wave scattering amplitude of the nuclear scattering in the presence of the Coulomb interaction is 
\begin{equation}
    f^{NC}_l(k,\theta)=(2l+1)\frac{1}{2ik}e^{2i\delta^{C}_{l}}(e^{2i\delta^{NC}_l}-1)P_l(\cos\theta)
    \label{eq:amp_orgin}
\end{equation}
$\delta^{NC}_l$ is the nuclear scattering phase shift with Coulomb interaction. It is worth noting that it is not identical to the pure nuclear scattering phase shift $\delta_l$ which is produced by $V_N(r)$ alone.
When $\delta^{NC}_l=0$ ($V_N(r)=0$, $\psi=\psi_{in}$), the correlation function is 
\begin{equation}
    C_{\rm{coul}}(k)=A_c(\eta)\int d^3r S(r) \vert F(-i\eta,1,ikr-i\vec{k}\cdot\vec{r})\vert^{2}
\end{equation}
Similarly, $\psi_{in}$ in Eq.~\ref{eq:coulomb_wave} can also be expanded into the eigen wave functions of angular momentum:
\begin{widetext}
\begin{equation}
    \sqrt{A_c(\eta)}e^{i\delta^{C}_{0}}e^{i\vec{k}\cdot \vec{r}}F(-i\eta,1,ikr-i\vec{k}\cdot\vec{r})=\sum_{l=0}^{\infty}(2l+1)P_l(\cos{\theta})i^{l}e^{i\delta^{C}_{l}}\frac{F_l(kr,\eta)}{kr}
\end{equation}
\end{widetext}
When $\delta^{NC}_l \neq 0$, a singularity appears in $e^{2i\delta^{C}_l}$ as $k \rightarrow 0$. So Eq.~\ref{eq:amp_orgin} is usually not directly used to calculate correlation function.
A renormalized scattering amplitude that cancels out the singularities has been employed \cite{Hamilton1973,Orlov:2020qpf}.
\begin{equation}
    \tilde{f_l}=\frac{1}{C_l^{2}(\eta)k(\cot\delta^{NC}_l-i)} 
    \label{eq:f_renomal_3}
\end{equation}
Among them, the factor $(2l+1)P_l(\cos{\theta})$ will only contribute a constant after integrating the KP equation, so it is omitted. Where
\begin{equation}
    C_0^2(\eta)=\frac{2\pi\eta}{e^{2\pi\eta}-1}
\end{equation}
\begin{equation}
    C_l^2(\eta)=C_{l-1}^2(\eta)(1+\frac{\eta^2}{l^2})
\end{equation}

Again the correlation function deviates from $C_{\rm{coul}}(k)$:
\begin{equation}
    C(k)=C_{\rm{coul}}(k)+\sum_{l=0}^{\infty}\Delta C_l(k)
\end{equation}
Using the orthogonality property of the Coulomb function, $\Delta C_l(k)$ can be calculated by the following integral
\begin{equation}
    \begin{aligned}
    &\quad\Delta C_l(k)\\
    &=\int d^3r S(r) \vert\sqrt{A_c(\eta)}[ e^{i\vec{k}\cdot \vec{r}}F(-i\eta,1,ikr-i\vec{k}\cdot\vec{r}) \\
    &+ \tilde{f_l} \frac{\tilde{G_{l}}(\rho,\eta)}{r}] \vert^{2} - C_{\rm{coul}}(k)
    \label{eq:delta_Ckstar}
    \end{aligned}
\end{equation}
Define a function $K_l(k^2)$ \cite{Alley:1993zz,Hamilton1973,Orlov:2020qpf}
\begin{equation}
    K_l(k^2)=k^{2l+1}\left[C_l^2(\eta)(\cot\delta_l^{NC}-i)+ 2\eta h(\eta)\prod_{n=1}^{l}\left(1+\frac{\eta^2}{n^2}\right)\right]
\end{equation}
where the $h(\eta)$ function is
\begin{equation}
\begin{aligned}    
    h(\eta)&=\Psi(i\eta)+(2i\eta)^{-1}-\ln(i\eta)\\
    &=\frac{iC_0^2(\eta)}{2\eta}+\eta^2\sum_{n=1}^{\infty}\frac{1}{n(n^2+\eta^2)}-\ln\eta-\gamma
\end{aligned}
\end{equation}
$\Psi$ is the digamma function. $\gamma=0.5772157...$ is the Euler constant.
Then Eq.\ref{eq:f_renomal_3} becomes
\begin{equation}
    \tilde{f}_{l}=\frac{k^{2l}}{K_l(k^2)-2\eta k^{2l+1} h(\eta) \prod_{n=1}^{l}\left(1+\frac{\eta^2}{n^2}\right)}
\end{equation}
The $K_l(k^2)$ cancels out the divergent terms, and it can be expanded as a Taylor series of even powers. This is the well-known effective range expansion
\begin{equation}
    \Re{K_l(k^2)}=\sum_{n=0}a^l_nk^{2n}
    \label{eq:eff_Range_expand}
\end{equation}
Specifically, when $l = 0$, retaining the quadratic term and performing the rearrangement leads to the widely used s-wave scattering amplitude with Coulomb correction:
\begin{equation}
    \tilde{f_0}=\left[-\frac{1}{a_0}+\frac{1}{2}r_0k^2-2k\eta \Re{h(\eta)} - ikC_0^2(\eta)\right]^{-1}
    \label{eq:effR_C}
\end{equation}
When $\eta\rightarrow0$, Eq.\ref{eq:effR_C} reduces to Eq.\ref{eq:effR_noC}.

\section{Numerical Calculation of $^{4}\rm{Li}$ and $^{5}\rm{Li}$}

\subsection{Sampling Algorithm}
For a static Gaussian source given by Eq.~\ref{eq:gauss_source}, the correlation function can be obtained by direct radial integration in the pair rest frame. However, to account for the momentum-space phase-space distribution and the Lorentz contraction of the emission source due to the pair center-of-mass motion, we adopt a Monte Carlo sampling procedure\cite{Xi:2019vev}. In this paper, for simplicity, we set the phase space distribution as a Boltzmann distribution
\begin{equation}
    \frac{d^3N}{2\pi m_{T}dm_{T}dyd\phi}\propto m_{T}e^{-\frac{m_{T}}{T}}.
    \label{eq:phase_space}
\end{equation}
where $m_{T}=\sqrt{m^2+p_{T}^2}$ is the transverse mass of the particle. Here, we set $T=155$ MeV. In this procedure, the relative distance $\vec{r}$ is sampled according to the spherical Gaussian distribution in the pair rest frame. The single-particle momenta $\vec{p}_1$ and $\vec{p}_2$ are sampled according to the Boltzmann distribution of Eq.~\ref{eq:phase_space}, from which the pair center-of-mass velocity $\vec{\beta}$ and the relative momentum $k$ are determined. The vector $\vec{r}$ is then boosted to $\vec{r}'$ by $\vec{\beta}$ to evaluate the Lorentz-contracted source shape in the laboratory frame. 

It is important to emphasize that the partial-wave decomposition and the asymptotic wave functions are evaluated in the pair rest frame, where spherical symmetry is preserved. The boost is applied only to the sampled coordinates for the purpose of constructing the correlation function in the laboratory frame. This procedure follows the standard LL framework and has been validated in previous femtoscopic analyses \cite{STAR:2024zvj,STAR:2025jwe}.

Noticing that the source function Eq.\ref{eq:gauss_source} is normalized, it is straightforward to see that
\begin{equation}
    \Delta C_l(k)=\frac{\int d^3r S(r) w_l(\vec{r'},\vec{k})}{\int d^3r S(r)}=\langle w_l(\vec{r'},\vec{k}) \rangle_{r'}
\end{equation}
where $w_l(\vec{r'},\vec{k})$ is the integration kernel from Eq.\ref{eq:delta_Ckstar}:
\begin{equation}
\begin{aligned}
    &\quad w_l(\vec{r'},\vec{k})\\
    &=\vert\sqrt{A_c(\eta)}[ e^{i\vec{k}\cdot \vec{r}}F(-i\eta,1,ikr-i\vec{k}\cdot\vec{r}) + \tilde{f_l} \frac{\tilde{G_{l}}(\rho,\eta)}{r}] \vert^{2} \\ 
    &-\vert\sqrt{A_c(\eta)}[ e^{i\vec{k}\cdot \vec{r}}F(-i\eta,1,ikr-i\vec{k}\cdot\vec{r}) \vert^{2}
\end{aligned}
\end{equation}
The notation $\langle \rangle_{r'}$ indicates taking the average over all random samples.

\subsection{LL model with $r_c$ cutoff}
It is noted that when $k$ is finite and $r\rightarrow 0$, $G_l \sim r^{-l}$. In Eq.\ref{eq:coulomb_wave}, $\vert\psi_{out}\vert^{2}\sim r^{-2(l+1)}$. This divergent behavior is expected. Based on the short-range nature of the nuclear force, we use the asymptotic wave function instead of the exact wave function in the region where the nuclear force vanishes. Obviously, the asymptotic wave function cannot be used within the range of nuclear force.

To avoid the divergence of the wave function at $r\rightarrow 0$, a cutoff is applied:
\begin{equation}
    \psi=\left\{
    \begin{aligned}
        &\psi_{in} &\qquad (r<r_c)\\
        &\psi_{in}+\sum_l f \cdot \psi_{out}^{l} &\qquad (r>r_c)
    \end{aligned}
    \right.
    \label{eq:psi_cutoff}
\end{equation}

The cutoff parameter $r_c$ introduced in Eq.\ref{eq:psi_cutoff} comes from the regularization scheme proposed by \cite{Murase:2025nlo} for the generalized LL formula. It used the asymptotic scattering wave function only for $r>r_c$. Since the interior wave function is not fixed by the phase shifts data alone, the cutoff dependence is treated as a model uncertainty and is examined by varying $r_c$. In the present study, we adopt a similar prescription and set $r_c = 1$ fm, which is roughly equivalent to the range of nuclear forces. In some studies of heavy-ion collision correlation functions \cite{STAR:2024zvj,STAR:2025jwe,STAR:2015kha}, considering only the s-wave contribution, the asymptotic wave function describes the experimental data well, and the obtained $R_g$ is usually larger than 1 fm, indicating that the cutoff is an appropriate approximation. In the $pp$ collision system, since it is impossible to ensure $R_g > r_c$, tools such as CATS \cite{Mihaylov:2018rva} have to be used to calculate the parameterized exact wave function.

\subsection{Phase Shift Data and Combination Coefficient} 
In this paper, the correlation functions of $p$-$^3$He and $p$-$^4\rm{He}$ pairs are calculated. The phase shift data are provided by low-energy nuclear scattering experiments $^3\rm{He}(p,p)^3\rm{He}$~\cite{Alley:1993zz} and $^4\rm{He}(p,p)^4\rm{He}$~\cite{Kumar:2022cdj}. The spins of $p$ and $^3$He are 1/2, while the spin of $^4\rm{He}$ is 0. Therefore, for the $p$-$^3$He system, the possible wave states to consider are $^{1}S_{0}$, $^{3}S_{1}$, $^{1}P_{1}$, $^{3}P_{0}$, $^{3}P_{1}$, and $^{3}P_{2}$. For the $p$-$^4\rm{He}$ system, the relevant states are $^{2}S_{1/2}$, $P_{1/2}$, and $P_{3/2}$. The $^{3}P_{2}$ channel excess of $p$-$^3$He corresponds to the ground state of $^{4}$Li, and the $P_{3/2}$ channel excess of $p$-$^4\rm{He}$ corresponds to the ground state of $^{5}$Li. The higher-order partial wave ($l>1$) phase shifts are only a few degrees in the energy range of interest, and are therefore neglected. The mixing between $^{1}P_{1}$ and $^{3}P_{1}$ ($J^\pi=1^-$) is also neglected.

\begin{figure}[htb]
\includegraphics[width=0.9\linewidth]{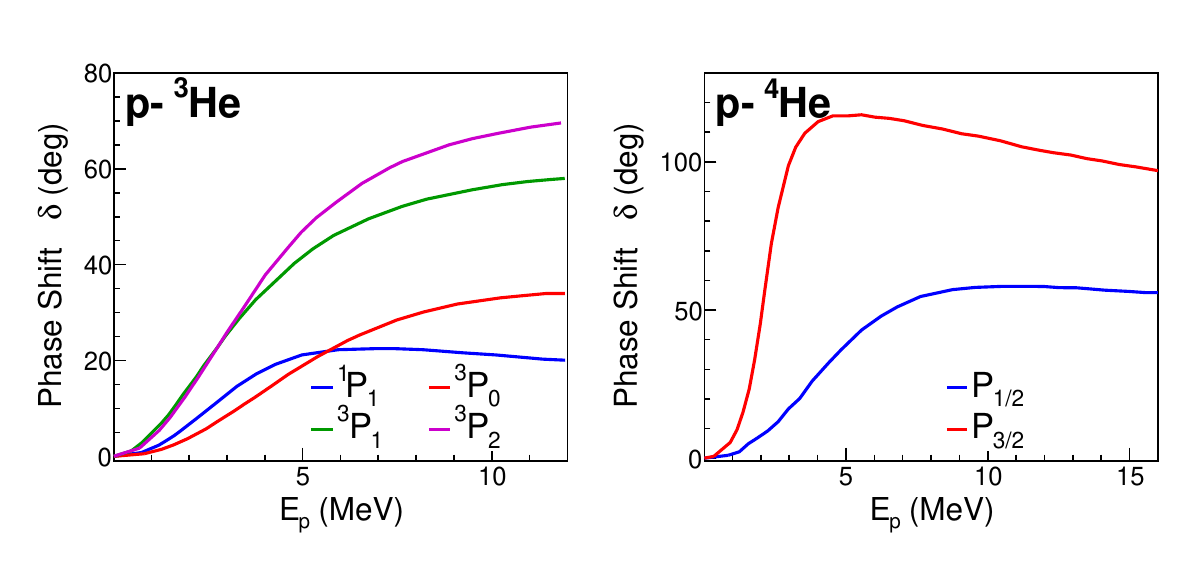}
\caption{The p-wave phase shift of $p$-$^3$He (left) and $p$-$^4\rm{He}$ (right). The data points are digitized from Ref.~\cite{Alley:1993zz,Kumar:2022cdj}. 
Only the linear interpolation result are shown in the figure. The horizontal axis $E_p$ represents the proton beam energy in the low-energy scattering experiment. In the non-relativistic case, $E_p=k^2(m_1+m_2)/2m_{red}m_2$, where $m_1$ is the mass of the proton, and $m_2$ is the mass of the target particle ($^3$He or $^4\rm{He}$). }
\label{fig:p_wave_PS}
\end{figure}

When $k\rightarrow0$, one should be careful when performing numerical calculations on Eq.\ref{eq:f_renomal_3} using phase-shift data. To avoid unnecessary divergence in the numerical calculation, we still apply Eq.\ref{eq:effR_C} to handle the s-wave.
By fitting the s-wave phase shift data with Eq.\ref{eq:eff_Range_expand}, we obtain 
\begin{align}
    a^s_0=7.16\rm{ fm} \qquad &r^{s}_0=2.75\rm{ fm} \qquad (^{1}S_{0}, \rm{singlet})\\
    a^t_0=9.07\rm{ fm} \qquad &r^{t}_0=1.47\rm{ fm} \qquad (^{3}S_{1}, \rm{triplet})
\end{align}
for the $p$-$^3$He system s-wave, and
\begin{equation}
    a_0=4.91\rm{ fm} \qquad r_{0}=1.39\rm{ fm} 
\end{equation}
for the $p$-$^4\rm{He}$ system s-wave. For the p-wave, we directly use the interpolated phase-shift data, which are shown in Fig.~\ref{fig:p_wave_PS}. For phase-shift data beyond the energy range shown in Fig.~\ref{fig:p_wave_PS}, the values are frozen, as they change slowly at higher energies.

After obtaining all the partial-wave components of the correlation function, the total correlation function can be constructed by the following formula
\begin{equation}
    C(k)=C_{\rm{coul}}(k)+\sum_{s}g_{\rm{spin}}\Delta C(s;k)+ g_p \sum_{s} g_{\rm{spin}}\Delta C(s;k)
\end{equation}
The summation index $s$ denotes the spin states. For the $p$-$^3$He system:
\begin{widetext}
\begin{equation}
\begin{aligned}
    C(k)&=C_{\rm{coul}}(k)+\left(\frac{1}{4}\Delta C(^{1}S_{0};k) + \frac{3}{4}\Delta C(^{3}S_{1};k)\right) \\
    &+ g_p\left(\frac{3}{12}\Delta C(^{1}P_{1};k) + \frac{1}{12}\Delta C(^{3}P_{0};k) +\frac{3}{12}\Delta C(^{3}P_{1};k) + \frac{5}{12}\Delta C(^{3}P_{2};k)\right)
    \label{eq:Ckstar_phe3}
    \end{aligned}
\end{equation}
and for the $p$-$^4\rm{He}$ system:
\begin{equation}
    C(k)=C_{\rm{coul}}(k)+\Delta C(^{2}S_{1/2};k) + g_p\left(\frac{1}{3}\Delta C(P_{1/2};k)+ \frac{2}{3}\Delta C(P_{3/2};k)\right)
    \label{eq:Ckstar_phe4}
\end{equation}
\end{widetext}
The numbers before each component are the spin degeneracy factors $g_{\rm{spin}}$. We assume unpolarized production for all particles, which is a common assumption in correlation function analysis. 
The spin degeneracy factor can be calculated by 
\begin{equation}
    g_{\rm{spin}} =  \frac{(2S+1)}{(2S_1+1)(2S_2+1)}\cdot\frac{(2J+1)}{(2L+1)(2S+1)}.
    \label{eq:g_s}
\end{equation}
We introduce a second fitting parameter $g_p$ as a scaling factor for the p-wave contributions in each system. The nuisance parameter $g_p$ is introduced to account for the residual short-range source-model dependence as well as the constants appearing in the scattering angle $\theta$ integral. In the Lednick\'y--Lyuboshitz framework, the exact wave function for $r<r_c$ is not solved, and the source function is approximated by a spherical Gaussian. The parameter $g_p$ effectively absorbs the systematic uncertainty arising from these short-range approximations for the p-wave contributions. The relative weights among different p-wave channels, however, remain strictly determined by spin degeneracy $g_{\rm{spin}}$ and phase-shift data. The uncertainty in the choice of $r_c$ can be treated as a systematic error in the fitting procedure.

\subsection{Results} 
\begin{figure}[htb]
\includegraphics[width=0.9\textwidth]{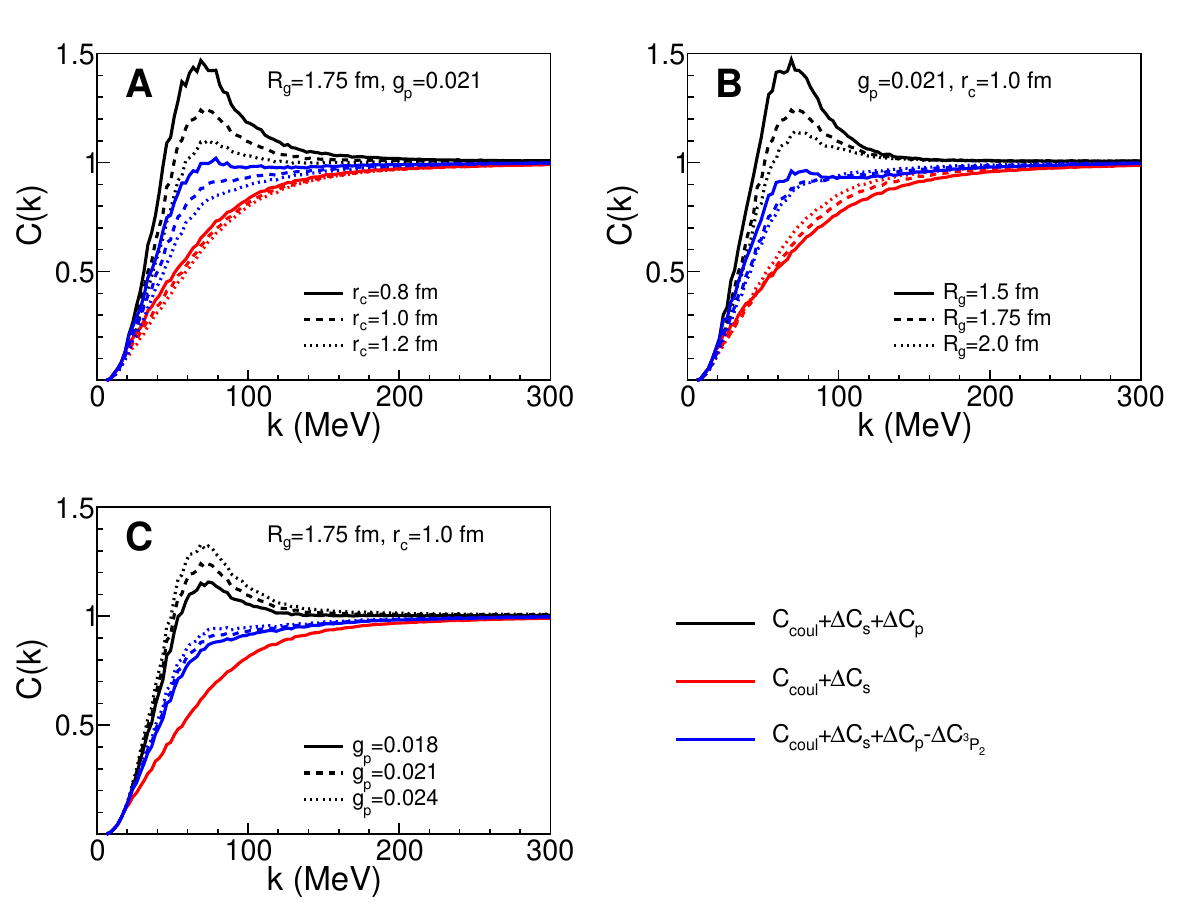}
\caption{Correlation functions for the $p$-$^3$He system. Panel (A): Dependence on the cutoff parameter $r_c$ for s-wave (red lines) , p-wave (black lines) and p-wave without resonance peak (blue lines). Panel (B): Dependence on the $R_{g}$ parameter. Panel (C): Dependence on the $g_{p}$ parameter.}
\label{fig2}
\end{figure}

Figure \ref{fig2} shows the calculated correlation functions for the $p$-$^3$He system. We present the results in three scenarios: (i) Coulomb baseline plus $s$-wave contributions only, which corresponds to the traditional LL model; (ii) Coulomb baseline plus all $s$-wave and $p$-wave contributions, which is the full calculation to be compared with experimental data; and (iii) Coulomb baseline plus all $s$-wave and non-resonant $p$-wave contributions (i.e., excluding the resonant $^3P_2$ channel), which represents the non-resonant component background for the $^4$Li(g.s.) signal.

As shown in Fig.~\ref{fig2}, the calculation naturally exhibits a peak at $k \approx 72$ MeV/c, which appears near the expected resonance energy region of $^4$Li. Even with the resonance peak, the correlation function remains below unity in the low-momentum region due to the repulsive Coulomb and $s$-wave interactions. This suggests that without proper background subtraction, the extracted signal number in this kinematic region would be negative. The results also show that the contributions of other P-wave components cannot be neglected. Under the assumption of unpolarized particles, the relative proportions of the four $p$-wave contributions, as well as the position and width of the peak, are entirely determined by the phase shifts and the spin degeneracy factors $g_{\rm{spin}}$. It is expected that due to the finite detector resolution, the resonance peaks in experimental data might be slightly wider than the calculated results, but the position of the resonance peak is still expected to change very little.

The cutoff $r_c$ is roughly set to the range of nuclear forces. However, there is no clear method to determine its exact value. It is not a physical parameter and therefore need not be included in the fitting process. A practical approach is to fix $r_c = 1$ fm and then check the sensitivity of the correlation function to this choice. Panel (A) of Fig.~\ref{fig2} shows that $r_c$ has a small effect on the s-wave but a significant impact on the magnitude of the p-wave. Therefore, as mentioned earlier, the $g_p$ factor needs to be introduced to eliminate the arbitrariness of the $r_c$ selection. 

We also check the dependence of the correlation function on the two fitting parameters. It is shown that the larger the source size $R_g$, the closer the correlation function is to unity. The $g_p$ factor only affects the amplitude of the p-wave contribution.

\begin{figure}[htb]
    \centering
    \includegraphics[width=0.9\textwidth]{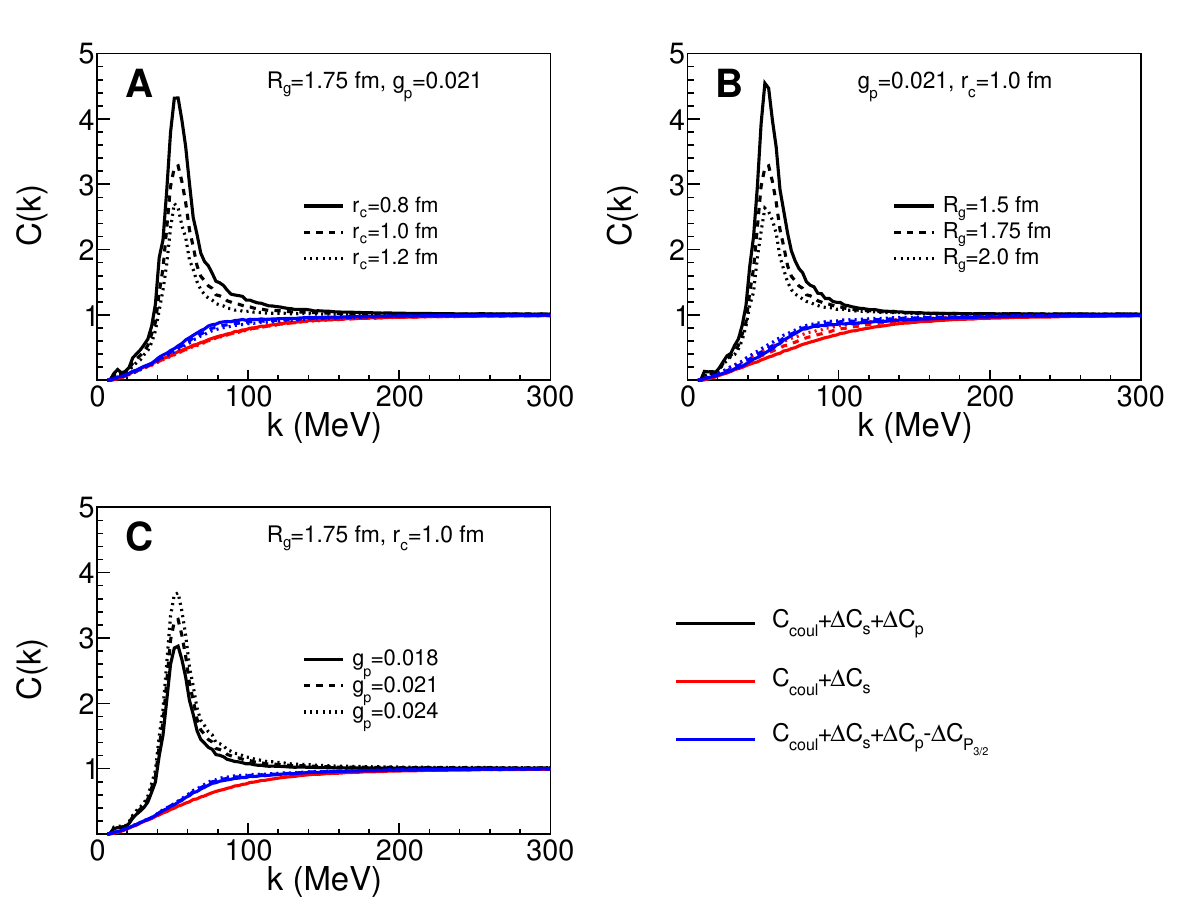}
    \caption{Correlation functions for the $p$-$^4$He system. The calculation naturally appear a peak at $k \approx 50$ MeV/c, which is the resonance of $^5$Li. Panel (A): Dependence on the cutoff parameter $r_c$ for s-wave (red lines) , p-wave (black lines) and p-wave without resonance peak (blue lines). Panel (B): Dependence on the $R_{g}$ parameter. Panel (C): Dependence on the $g_{p}$ parameter.}
    \label{fig3}
\end{figure}

Similarly, we also calculated the correlation function of the $p$-$^4\rm{He}$ system, which is shown in Fig.~\ref{fig3}. As shown in Fig.~\ref{fig:p_wave_PS}, the $P_{3/2}$ ($^5\rm{Li}$) partial wave phase shift of the $p$-$^4\rm{He}$ system passing through 90 degrees indicates that this resonance is stronger than that of $^4\rm{Li}$. Therefore, under the same $R_g$ and $g_p$ parameters, the resonance peak is significantly higher and narrower than that of $^4\rm{Li}$. 

\begin{widetext}
\begin{figure*}[htb]
    \centering
    \includegraphics[width=1.0\columnwidth]{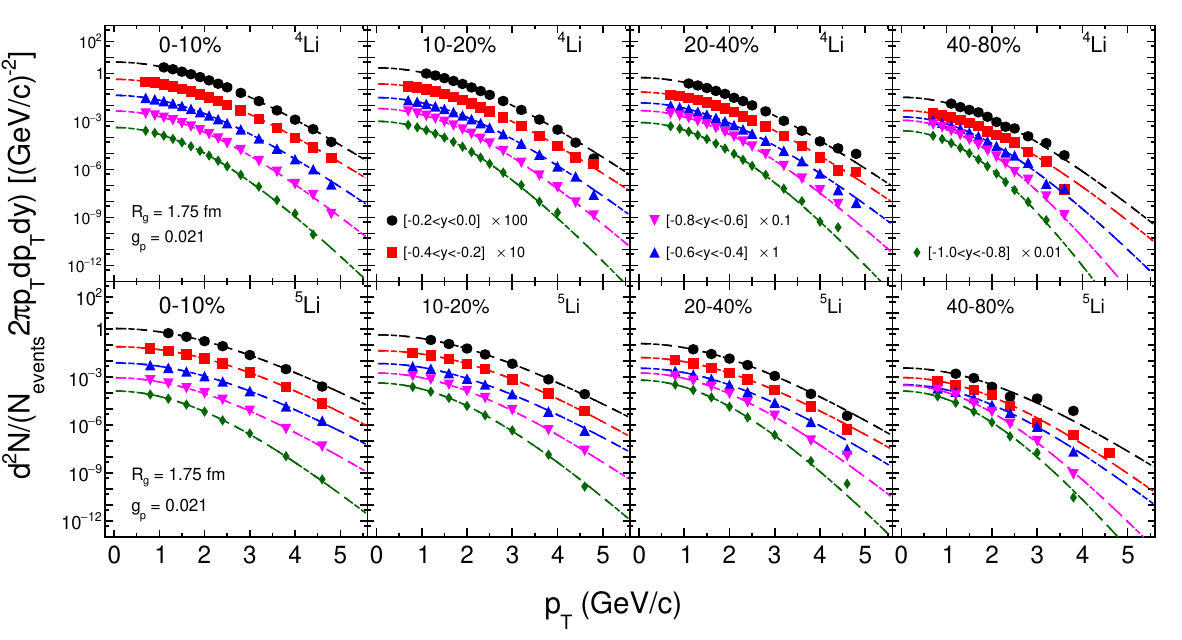}
    \caption{Transverse momentum spectra ($p_{T}$) of $^{4}\rm{Li}$ (top) and $^{5}\rm{Li}$ (bottom) from different rapidity ranges and centrality bins. The proton, $^{3}\rm{He}$ and $^{4}\rm{He}$ spectra used in the calculations were obtained from the STAR measurement in Au+Au collisions at $\sqrt{s_{NN}}=3$ GeV\cite{STAR:2023uxk}. The parameters $R_g$ and $g_p$ are fixed at 1.75 fm and 0.021. The dashed lines are blast-wave function fits.}
    \label{fig4}
\end{figure*}
\end{widetext}

\begin{figure}
    \centering
    \includegraphics[width=0.9\textwidth]{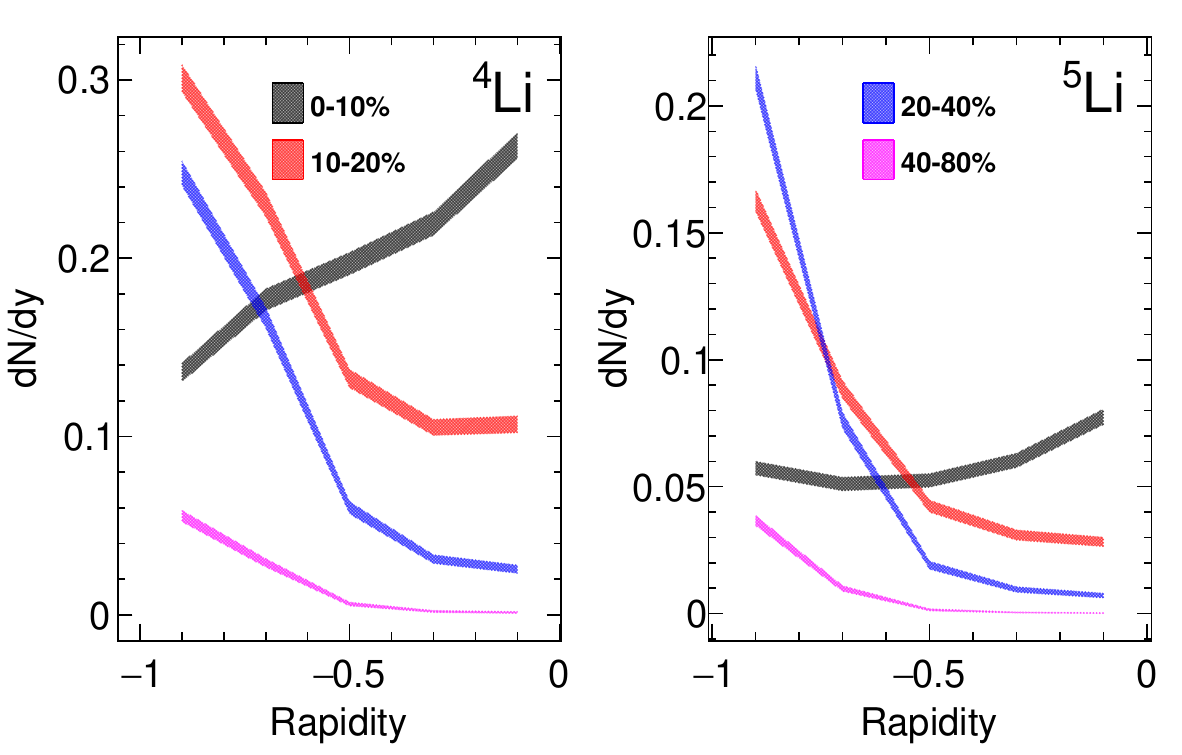}
    \caption{The rapidity dependence of $dN/dy$ for $^{4}\rm{Li}$ (left) and $^{5}\rm{Li}$ (right) in different centrality bins. The parameters $R_g$ and $g_p$ are fixed at 1.75 fm and 0.021.}
    \label{fig5}
\end{figure}

Once the fit parameters $R_g$ and $g_p$ are obtained by fitting the $p$-$^3\rm{He}$ correlation function, the $^{3}P_{2}$ channel excess can be extracted. Then the $^{4}\rm{Li}$ spectra can be calculated based on the spectra of $p$ and $^3$He. Following the definition of the correlation function
\begin{equation}
    \frac{dN_{pair}^{6}}{d^{3}p_{1}\,d^{3}p_{2}}=C(k)\frac{dN_{1}^{3}}{d^{3}p_{1}}\frac{dN_{2}^{3}}{d^{3}p_{2}},
\end{equation}
where 1 and 2 represent the two particles, respectively. 
It should be noted that the $^3P_2$ partial-wave amplitude generally contains both resonant and non-resonant contributions, along with their interference. In the range of the resonance peak, the $\Delta C(^3P_2)$ term serves as an effective contribution for the resonance pair correlation. Therefore, the corresponding resonance yield can be practically estimated by integrating this partial-wave excess over the relevant momentum range, as shown in Eq.\ref{eq:N_Li4}. According to Eq.\ref{eq:Ckstar_phe3}, the contributed by the $^3P_2$ channel excess is
\begin{equation}
    \frac{dN_{^{4}\rm{Li}}^{6}}{d^{3}P\,d^{3}k}=\frac{5g_p}{12}\Delta C(^{3}P_{2};k)\frac{dN_{1}^{3}}{d^{3}p_{1}}\frac{dN_{2}^{3}}{d^{3}p_{2}}
    \label{eq:N_Li4}
\end{equation}
where $\vec{P} = \vec{p}_1 + \vec{p}_{2}$ is the momentum of pair. According to Fig.~\ref{fig2}, $\frac{5g_p}{12}\Delta C(^{3}P_{2};k)$ is nonzero within a range of approximately 200 MeV. Within this range, it can be assumed that the production yields of $p$ and $^3$He vary slowly. Integrating Eq.\ref{eq:N_Li4} over $k$ gives
\begin{equation}
    \frac{dN_{^{4}\rm{Li}}^{3}}{d^{3}P} \approx I(^{3}P_{2})\frac{dN_{1}^{3}}{d^{3}p_{1}}\frac{dN_{2}^{3}}{d^{3}p_{2}}|_{k=0}
    \label{Li4_N}
\end{equation}
where
\begin{equation}
    I \approx \int_0^{\rm{200 MeV}}\frac{5g_p}{12}\Delta C(^{3}P_{2};k)4\pi k{}^2dk.
\end{equation}
When $k=0$, we have $\frac{1}{3}p_{2}\approx p_{1}$. Eq.\ref{eq:N_Li4} can be used to calculate the $p_T$ spectra of the resonance. Fig.~\ref{fig4} shows the transverse momentum spectra for $^{4}\rm{Li}$ and $^{5}\rm{Li}$ in 0-10\%, 10-20\%, 20-40\%, and 40-80\% centrality. The proton, $^{3}\rm{He}$ and $^{4}\rm{He}$ spectra use the STAR measurement in Au+Au collisions at $\sqrt{s_{NN}}=3$ GeV\cite{STAR:2023uxk}. The results are shown for rapidity intervals of width 0.2. Fig.~\ref{fig5} shows the rapidity dependence of integrated yield ($dN/dy$) for $^{4}\rm{Li}$ and $^{5}\rm{Li}$. Due to baryon stopping and spectator effects, $dN/dy$ for $^{4}\rm{Li}$ and $^{5}\rm{Li}$ decreases from mid-rapidity toward target rapidity in central collisions, while it increases in peripheral collisions. This behavior is inherited from the $dN/dy$ rapidity dependence of the $^{3}\rm{He}$ and $^{4}\rm{He}$ spectra used as input. We emphasize that Fig.~\ref{fig4} and Fig.~\ref{fig5} are  demonstration with fix parameters rather than quantitative predictions.

According to Eq.\ref{Li4_N}, it is easy to obtain that the particle yield ratio $N_{^{4}\rm{Li}}/(N_{p}\times N_{^{3}\rm{He}})$ or $N_{^{5}\rm{Li}}/(N_{p}\times N_{^{4}\rm{He}})$ is completely determined by the contribution to the corresponding partial wave excess in the correlation function. With the currently same parameter configuration, it is calculated that the double ratio $N_{^{4}\rm{Li}}N_{^{4}\rm{He}}/N_{^{5}\rm{Li}}N_{^{3}\rm{He}}\approx 0.6$, which is caused by the difference of resonance strength. This indicates that the resonance strength is positively correlated with its yield.

\section{Summary and Outlook}
In this paper, we have developed a partial-wave method based on the Lednicky-Lyuboshitz model to extract the resonance signals of unstable light nuclei from two-particle correlation functions in relativistic heavy-ion collisions. The key feature of our approach is the extension of the LL model originally limited to $s$-wave interactions to higher order partial waves, enabling a comprehensive treatment of final-state interactions where the resonant states of interest often reside in higher angular momentum channels.

The theoretical framework combines the Koonin-Pratt equation with an asymptotic wave function that incorporates both nuclear and Coulomb interactions. By directly using phase-shift data from low-energy nuclear scattering experiments as input, our method avoids the need for detailed and often model-dependent nuclear potential parameterizations, employing a Coulomb-distorted asymptotic treatment through a renormalized scattering amplitude. 

As a demonstration, we have applied our method to the specific cases of $^4$Li and $^5$Li. For the $p$-$^3$He system ($^4$Li, $^3P_2$ resonance), we have included the contributions from all relevant partial waves: $^1S_0$, $^3S_1$, $^1P_1$, $^3P_0$, $^3P_1$, and the resonant $^3P_2$ channel. For the $p$-$^4\rm{He}$ system ($^5$Li, $P_{3/2}$ resonance), the considered partial waves include $^2S_{1/2}$, $P_{1/2}$, and the resonant $P_{3/2}$ channel. The phase-shift data are derived from low-energy nuclear scattering experiments and naturally contain information on resonant states. The total correlation function is constructed as a spin-weighted sum of all partial-wave contributions. In total, only two parameters $R_g$ and $g_p$ need to be determined through experimental data.

The method's applicability is based on the availability of reliable phase-shift data and the assumption of a Gaussian source. In the few-GeV collision energy range, a large number of light nuclei are expected to be produced. We suggest that experimental collaborations apply this framework to existing HIC data from STAR or CEE.  Our numerical results demonstrate that the resonant signals can be effectively separated from the non-resonant background, providing a practical tool for the experimental study of unstable light nuclei in heavy-ion collisions. The method can be applied to a broader range of unstable nuclei, such as $^6\rm{He}$ (16.76 MeV), $^6\rm{Li}$ (2.18 MeV), $^7\rm{Li}$ (4.63 MeV), $^8\rm{Be}$ (g.s.), etc. By measuring these unstable light nucleus resonant states, it is possible to constrain the theoretical models of the unstable light nucleus production process in relativistic heavy-ion collisions. 

\begin{acknowledgments}
The authors thank helpful discussions with Jiangyong Jia, Kai-Jia Sun, Yun-Peng Zheng, Liang Zhang, Yixuan Zhao. This work is partially supported by the Postdoctoral Fellowship Program of China Postdoctoral Science Foundation (GZC20252245).
\end{acknowledgments}

\appendix
\bibliography{mainbib}

@article{STAR:2019sjh,
    author = "Adam, Jaroslav and others",
    collaboration = "STAR",
    title = "{Beam energy dependence of (anti-)deuteron production in Au + Au collisions at the BNL Relativistic Heavy Ion Collider}",
    eprint = "1903.11778",
    archivePrefix = "arXiv",
    primaryClass = "nucl-ex",
    doi = "10.1103/PhysRevC.99.064905",
    journal = "Phys. Rev. C",
    volume = "99",
    number = "6",
    pages = "064905",
    year = "2019"
}

@article{STAR:2023uxk,
    author = "Abdulhamid, Muhammad and others",
    collaboration = "STAR",
    title = "{Production of protons and light nuclei in Au+Au collisions at sNN=3 GeV with the STAR detector}",
    eprint = "2311.11020",
    archivePrefix = "arXiv",
    primaryClass = "nucl-ex",
    doi = "10.1103/PhysRevC.110.054911",
    journal = "Phys. Rev. C",
    volume = "110",
    number = "5",
    pages = "054911",
    year = "2024"
}

@article{STAR:2022hbp,
    author = "Abdulhamid, Muhammad and others",
    collaboration = "STAR",
    title = "{Beam Energy Dependence of Triton Production and Yield Ratio ($\mathrm{N}_t \times \mathrm{N}_p/\mathrm{N}_d^2$) in Au+Au Collisions at RHIC}",
    eprint = "2209.08058",
    archivePrefix = "arXiv",
    primaryClass = "nucl-ex",
    doi = "10.1103/PhysRevLett.130.202301",
    journal = "Phys. Rev. Lett.",
    volume = "130",
    pages = "202301",
    year = "2023"
}

@article{Jin:2025biy,
    author = "Jin, Yixuan",
    title = "{Production of light nuclei in Au+Au collisions at {\ensuremath{\sqrt{}}}sNN = 7.7{\ensuremath{-}}27 GeV from STAR BES-II}",
    eprint = "2512.05295",
    archivePrefix = "arXiv",
    primaryClass = "nucl-ex",
    doi = "10.1051/epjconf/202636402009",
    journal = "EPJ Web Conf.",
    volume = "364",
    pages = "02009",
    year = "2026"
}

@article{ALICE:2015wav,
    author = "Adam, Jaroslav and others",
    collaboration = "ALICE",
    title = "{Production of light nuclei and anti-nuclei in pp and Pb-Pb collisions at energies available at the CERN Large Hadron Collider}",
    eprint = "1506.08951",
    archivePrefix = "arXiv",
    primaryClass = "nucl-ex",
    reportNumber = "CERN-PH-EP-2015-025",
    doi = "10.1103/PhysRevC.93.024917",
    journal = "Phys. Rev. C",
    volume = "93",
    number = "2",
    pages = "024917",
    year = "2016"
}

@article{ALICE:2017xrp,
    author = "Acharya, Shreyasi and others",
    collaboration = "ALICE",
    title = "{Production of deuterons, tritons, $^{3}$He nuclei and their antinuclei in pp collisions at $\mathbf{\sqrt{{\textit s}}}$ = 0.9, 2.76 and 7 TeV}",
    eprint = "1709.08522",
    archivePrefix = "arXiv",
    primaryClass = "nucl-ex",
    reportNumber = "CERN-EP-2017-255",
    doi = "10.1103/PhysRevC.97.024615",
    journal = "Phys. Rev. C",
    volume = "97",
    number = "2",
    pages = "024615",
    year = "2018"
}

@article{ALICE:2017jmf,
    author = "Acharya, Shreyasi and others",
    collaboration = "ALICE",
    title = "{Production of $^{4}$He and $^{4}\overline{\textrm{He}}$ in Pb-Pb collisions at $\sqrt{s_{\mathrm{NN}}}$ = 2.76 TeV at the LHC}",
    eprint = "1710.07531",
    archivePrefix = "arXiv",
    primaryClass = "nucl-ex",
    reportNumber = "CERN-EP-2017-266",
    doi = "10.1016/j.nuclphysa.2017.12.004",
    journal = "Nucl. Phys. A",
    volume = "971",
    pages = "1--20",
    year = "2018"
}

@article{ALICE:2020foi,
    author = "Acharya, S. and others",
    collaboration = "ALICE",
    title = "{(Anti-)deuteron production in pp collisions at $\sqrt{s}=13 \ \text {TeV}$}",
    eprint = "2003.03184",
    archivePrefix = "arXiv",
    primaryClass = "nucl-ex",
    reportNumber = "CERN-EP-2020-025",
    doi = "10.1140/epjc/s10052-020-8256-4",
    journal = "Eur. Phys. J. C",
    volume = "80",
    number = "9",
    pages = "889",
    year = "2020"
}

@article{ALICE:2019bnp,
    author = "Acharya, Shreyasi and others",
    collaboration = "ALICE",
    title = "{Multiplicity dependence of light (anti-)nuclei production in p-Pb collisions at $\sqrt{s_{\rm{NN}}}$ = 5.02 TeV}",
    eprint = "1906.03136",
    archivePrefix = "arXiv",
    primaryClass = "nucl-ex",
    reportNumber = "CERN-EP-2019-120",
    doi = "10.1016/j.physletb.2019.135043",
    journal = "Phys. Lett. B",
    volume = "800",
    pages = "135043",
    year = "2020"
}

@article{ALICE:2021mfm,
    author = "Acharya, Shreyasi and others",
    collaboration = "ALICE",
    title = "{Production of light (anti)nuclei in pp collisions at $ \sqrt{s} $ = 13 TeV}",
    eprint = "2109.13026",
    archivePrefix = "arXiv",
    primaryClass = "nucl-ex",
    reportNumber = "CERN-EP-2021-194",
    doi = "10.1007/JHEP01(2022)106",
    journal = "JHEP",
    volume = "1",
    pages = "106",
    year = "2022"
}

@article{ALICE:2021ovi,
    author = "Acharya, Shreyasi and others",
    collaboration = "ALICE",
    title = "{Production of light (anti)nuclei in pp collisions at $\sqrt{s} = 5.02$~TeV}",
    eprint = "2112.00610",
    archivePrefix = "arXiv",
    primaryClass = "nucl-ex",
    reportNumber = "CERN-EP-2021-250",
    doi = "10.1140/epjc/s10052-022-10241-z",
    journal = "Eur. Phys. J. C",
    volume = "82",
    number = "4",
    pages = "289",
    year = "2022"
}

@article{ALICE:2025byl,
    author = "Acharya, S. and others",
    collaboration = "ALICE",
    title = "{Observation of deuteron and antideuteron formation from resonance-decay nucleons}",
    eprint = "2504.02393",
    archivePrefix = "arXiv",
    primaryClass = "nucl-ex",
    reportNumber = "CERN-EP-2025-081",
    doi = "10.1038/s41586-025-09775-5",
    journal = "Nature",
    volume = "648",
    number = "8093",
    pages = "306--311",
    year = "2025"
}

@article{Andronic:2010qu,
    author = "Andronic, A. and Braun-Munzinger, P. and Stachel, J. and Stocker, H.",
    title = "{Production of light nuclei, hypernuclei and their antiparticles in relativistic nuclear collisions}",
    eprint = "1010.2995",
    archivePrefix = "arXiv",
    primaryClass = "nucl-th",
    doi = "10.1016/j.physletb.2011.01.053",
    journal = "Phys. Lett. B",
    volume = "697",
    pages = "203--207",
    year = "2011"
}

@article{Bazak:2018hgl,
    author = "Bazak, Sylwia and Mrowczynski, Stanislaw",
    title = "{$^4{\rm He}$ vs. $^4{\rm Li}$ and production of light nuclei in relativistic heavy-ion collisions}",
    eprint = "1802.08212",
    archivePrefix = "arXiv",
    primaryClass = "nucl-th",
    doi = "10.1142/S0217732318501420",
    journal = "Mod. Phys. Lett. A",
    volume = "33",
    number = "25",
    pages = "1850142",
    year = "2018"
}

@article{ALICE:2025wuy,
    author = "Acharya, Shreyasi and others",
    collaboration = "ALICE",
    title = "{Femtoscopic study of the proton-proton and proton-deuteron systems in heavy-ion collisions at the LHC}",
    eprint = "2505.01061",
    archivePrefix = "arXiv",
    primaryClass = "nucl-ex",
    reportNumber = "CERN-EP-2025-096",
    doi = "10.1016/j.physletb.2025.139921",
    journal = "Phys. Lett. B",
    volume = "871",
    pages = "139921",
    year = "2025"
}

@article{STAR:2024zvj,
    author = "Aboona, B. E. and others",
    collaboration = "STAR",
    title = "{Light nuclei femtoscopy and baryon interactions in 3 GeV Au+Au collisions at RHIC}",
    eprint = "2410.03436",
    archivePrefix = "arXiv",
    primaryClass = "nucl-ex",
    doi = "10.1016/j.physletb.2025.139412",
    journal = "Phys. Lett. B",
    volume = "864",
    pages = "139412",
    year = "2025"
}

@article{Zheng:2025ngn,
    author = "Zheng, Yun-Peng and Liu, Dai-Neng and Chen, Lie-Wen and Chen, Jin-Hui and Ko, Che Ming and Ma, Yu-Gang and Sun, Kai-Jia and Xu, Jun and Zhou, Bo",
    title = "{Global spin alignment of (anti-)Li4 in noncentral heavy-ion collisions}",
    eprint = "2509.15286",
    archivePrefix = "arXiv",
    primaryClass = "nucl-th",
    doi = "10.1103/gb8k-5vg6",
    journal = "Phys. Rev. C",
    volume = "113",
    number = "5",
    pages = "054906",
    year = "2026"
}

@article{STAR:2011eej,
    author = "Agakishiev, H. and others",
    collaboration = "STAR",
    title = "{Observation of the antimatter helium-4 nucleus}",
    eprint = "1103.3312",
    archivePrefix = "arXiv",
    primaryClass = "nucl-ex",
    doi = "10.1038/nature10079",
    journal = "Nature",
    volume = "473",
    pages = "353",
    year = "2011",
    note = "[Erratum: Nature 475, 412 (2011)]"
}

@article{Xi:2019vev,
    author = "Xi, Bao-Shan and Zhang, Zheng-Qiao and Zhang, Song and Ma, Yu-Gang",
    title = "{Searching for $\overline{^4Li}$ via the momentum-correlation function of $\overline{p}-\overline{^3He}$}",
    eprint = "1909.03157",
    archivePrefix = "arXiv",
    primaryClass = "nucl-th",
    doi = "10.1103/PhysRevC.102.064901",
    journal = "Phys. Rev. C",
    volume = "102",
    number = "6",
    pages = "064901",
    year = "2020"
}

@article{Reichstein:1971wa,
    author = "Reichstein, I. and Thompson, D. R. and Tang, Y. C.",
    title = "{Study of p + He-3 and n + H-3 Systems with the Resonating-Group Method}",
    doi = "10.1103/PhysRevC.3.2139",
    journal = "Phys. Rev. C",
    volume = "3",
    pages = "2139--2148",
    year = "1971"
}

@article{Tombrello:1965zz,
    author = "Tombrello, T. A.",
    title = "{Phase-Shift Analysis for He-3 (p, p) He-3}",
    doi = "10.1103/PhysRev.138.B40",
    journal = "Phys. Rev.",
    volume = "138",
    pages = "B40--B47",
    year = "1965"
}

@article{Alley:1993zza,
    author = "Alley, M. T. and Knutson, L. D.",
    title = "{Spin correlation measurements for p- He-3 elastic scattering between 4.0 and 10.0 MeV}",
    doi = "10.1103/PhysRevC.48.1890",
    journal = "Phys. Rev. C",
    volume = "48",
    pages = "1890--1900",
    year = "1993"
}

@article{Alley:1993zz,
    author = "Alley, M. T. and Knutson, L. D.",
    title = "{Effective range parametrization of phase shifts for p- He-3 elastic scattering between 0 and 12 MeV}",
    doi = "10.1103/PhysRevC.48.1901",
    journal = "Phys. Rev. C",
    volume = "48",
    pages = "1901--1909",
    year = "1993"
}

@article{Lednicky:1981su,
    author = "Lednicky, R. and Lyuboshits, V. L.",
    title = "{Final State Interaction Effect on Pairing Correlations Between Particles with Small Relative Momenta}",
    reportNumber = "JINR-E2-81-453",
    journal = "Yad. Fiz.",
    volume = "35",
    pages = "1316--1330",
    year = "1981"
}

@article{Koonin:1977fh,
    author = "Koonin, S. E.",
    title = "{Proton Pictures of High-Energy Nuclear Collisions}",
    doi = "10.1016/0370-2693(77)90340-9",
    journal = "Phys. Lett. B",
    volume = "70",
    pages = "43--47",
    year = "1977"
}

@article{Viviani:1998gr,
    author = "Viviani, M. and Rosati, S. and Kievsky, A.",
    title = "{Neutron H-3 and proton He-3 zero energy scattering}",
    eprint = "nucl-th/9807059",
    archivePrefix = "arXiv",
    doi = "10.1103/PhysRevLett.81.1580",
    journal = "Phys. Rev. Lett.",
    volume = "81",
    pages = "1580--1583",
    year = "1998"
}

@article{Pudliner:1997ck,
    author = "Pudliner, B. S. and Pandharipande, V. R. and Carlson, J. and Pieper, Steven C. and Wiringa, Robert B.",
    title = "{Quantum Monte Carlo calculations of nuclei with A {\ensuremath{<}}= 7}",
    eprint = "nucl-th/9705009",
    archivePrefix = "arXiv",
    reportNumber = "ANL-PHY-8693-TH-97",
    doi = "10.1103/PhysRevC.56.1720",
    journal = "Phys. Rev. C",
    volume = "56",
    pages = "1720--1750",
    year = "1997"
}

@article{Wiringa:1994wb,
    author = "Wiringa, Robert B. and Stoks, V. G. J. and Schiavilla, R.",
    title = "{An Accurate nucleon-nucleon potential with charge independence breaking}",
    eprint = "nucl-th/9408016",
    archivePrefix = "arXiv",
    reportNumber = "PHY-7742-TH-94, CEBAF-TH-94-19",
    doi = "10.1103/PhysRevC.51.38",
    journal = "Phys. Rev. C",
    volume = "51",
    pages = "38--51",
    year = "1995"
}

@article{Mihaylov:2018rva,
    author = "Mihaylov, D. L. and Mantovani Sarti, V. and Arnold, O. W. and Fabbietti, L. and Hohlweger, B. and Mathis, A. M.",
    title = {{A femtoscopic Correlation Analysis Tool using the Schr{\"o}dinger equation (CATS)}},
    eprint = "1802.08481",
    archivePrefix = "arXiv",
    primaryClass = "hep-ph",
    doi = "10.1140/epjc/s10052-018-5859-0",
    journal = "Eur. Phys. J. C",
    volume = "78",
    number = "5",
    pages = "394",
    year = "2018"
}

@article{Ge:2025put,
    author = "Ge, Duo-Lun and Liu, Zhi-Wei and Lu, Jun-Xu and Geng, Li-Sheng",
    title = "{Deuteron-deuteron interaction and correlation function}",
    eprint = "2502.18872",
    archivePrefix = "arXiv",
    primaryClass = "nucl-th",
    doi = "10.1103/ttrc-qhv5",
    journal = "Phys. Rev. C",
    volume = "112",
    number = "3",
    pages = "034003",
    year = "2025"
}

@article{Hamilton1973,
    author  = "J.Hamilton and I. Overbo",
    title   = "Coulomb Corrections in Non-Relativistic Scattering",
    year    = "1973",
    journal = "Nuclear Physics",
    volume  = "B60",
    pages   = "443--477"
}

@article{Orlov:2020qpf,
    author = "Orlov, Yu. V.",
    title = "{The energies and ANCs for 5 Li resonances deduced from experimental p - {\ensuremath{\alpha}} scattering phase shifts using the effective-range and {\ensuremath{\Delta}} methods}",
    eprint = "2004.12855",
    archivePrefix = "arXiv",
    primaryClass = "nucl-th",
    doi = "10.1016/j.nuclphysa.2020.122060",
    journal = "Nucl. Phys. A",
    volume = "1004",
    pages = "122060",
    year = "2020"
}

@article{Murase:2025nlo,
    author = "Murase, Koichi and Hyodo, Tetsuo",
    title = "{Regularized Lednicky-Lyuboshitz formula for higher partial waves in femtoscopy}",
    eprint = "2509.22844",
    archivePrefix = "arXiv",
    primaryClass = "nucl-th",
    reportNumber = "RIKEN-iTHEMS-Report-25",
    month = "9",
    year = "2025",
    journal = "arXiv preprint"
}

@article{Kumar:2022cdj,
    author = "Kumar, Lalit and Awasthi, Shikha and Khachi, Anil and Sastri, O. S. K. S.",
    title = "{Phase Shift Analysis of Light Nucleon-Nucleus Elastic Scattering using Reference Potential Approach}",
    eprint = "2209.00951",
    archivePrefix = "arXiv",
    primaryClass = "nucl-th",
    month = "9",
    year = "2022",
    journal = "arXiv preprint"
}

@article{Dong:2020hxe,
    author = "Dong, Xiang-Kun and Guo, Feng-Kun and Zou, Bing-Song",
    title = "{Explaining the Many Threshold Structures in the Heavy-Quark Hadron Spectrum}",
    eprint = "2011.14517",
    archivePrefix = "arXiv",
    primaryClass = "hep-ph",
    doi = "10.1103/PhysRevLett.126.152001",
    journal = "Phys. Rev. Lett.",
    volume = "126",
    number = "15",
    pages = "152001",
    year = "2021"
}

@article{Zhang:2024qkg,
    author = "Zhang, Zhen-Hua and Guo, Feng-Kun",
    title = "{Classification of coupled-channel near-threshold structures}",
    eprint = "2407.10620",
    archivePrefix = "arXiv",
    primaryClass = "hep-ph",
    doi = "10.1016/j.physletb.2025.139387",
    journal = "Phys. Lett. B",
    volume = "863",
    pages = "139387",
    year = "2025"
}

@article{STAR:2025jwe,
    author = "Aboona, B. E. and others",
    collaboration = "STAR",
    title = "{First Observation of Deuteron-{\ensuremath{\Lambda}} Correlations at RHIC}",
    eprint = "2511.15493",
    archivePrefix = "arXiv",
    primaryClass = "nucl-ex",
    doi = "10.1103/3m26-5y83",
    journal = "Phys. Rev. Lett.",
    volume = "136",
    number = "24",
    pages = "242303",
    year = "2026"
}

@article{STAR:2015kha,
    author = "Adamczyk, L. and others",
    collaboration = "STAR",
    title = "{Measurement of Interaction between Antiprotons}",
    eprint = "1507.07158",
    archivePrefix = "arXiv",
    primaryClass = "nucl-ex",
    doi = "10.1038/nature15724",
    journal = "Nature",
    volume = "527",
    pages = "345--348",
    year = "2015"
}

@article{STAR:2018uho,
    author = "Adam, Jaroslav and others",
    collaboration = "STAR",
    title = "{The Proton-$\Omega$ correlation function in Au+Au collisions at $\sqrt{s_{NN}}$=200 GeV}",
    eprint = "1808.02511",
    archivePrefix = "arXiv",
    primaryClass = "hep-ex",
    doi = "10.1016/j.physletb.2019.01.055",
    journal = "Phys. Lett. B",
    volume = "790",
    pages = "490--497",
    year = "2019"
}

@article{STAR:2014dcy,
    author = "Adamczyk, L. and others",
    collaboration = "STAR",
    title = "{$\Lambda\Lambda$ Correlation Function in Au+Au collisions at $\sqrt{s_{NN}}=$ 200 GeV}",
    eprint = "1408.4360",
    archivePrefix = "arXiv",
    primaryClass = "nucl-ex",
    doi = "10.1103/PhysRevLett.114.022301",
    journal = "Phys. Rev. Lett.",
    volume = "114",
    number = "2",
    pages = "022301",
    year = "2015"
}

@article{Armstrong:2001mr,
    author = "Armstrong, T. A. and others",
    title = "{Production of particle unstable light nuclei in 11.5-A-GeV/C Au + Pt heavy ion collisions}",
    doi = "10.1103/PhysRevC.65.014906",
    journal = "Phys. Rev. C",
    volume = "65",
    pages = "014906",
    year = "2002"
}

@article{STAR:2017ckg,
    author = "Adamczyk, L. and others",
    collaboration = "STAR",
    title = "{Global $\Lambda$ hyperon polarization in nuclear collisions: evidence for the most vortical fluid}",
    eprint = "1701.06657",
    archivePrefix = "arXiv",
    primaryClass = "nucl-ex",
    doi = "10.1038/nature23004",
    journal = "Nature",
    volume = "548",
    pages = "62--65",
    year = "2017"
}

@article{STAR:2022fan,
    author = "Abdallah, M. S. and others",
    collaboration = "STAR",
    title = "{Pattern of global spin alignment of {\ensuremath{\phi}} and K$^{*0}$ mesons in heavy-ion collisions}",
    eprint = "2204.02302",
    archivePrefix = "arXiv",
    primaryClass = "hep-ph",
    doi = "10.1038/s41586-022-05557-5",
    journal = "Nature",
    volume = "614",
    number = "7947",
    pages = "244--248",
    year = "2023"
}

@article{Bazak:2020wjn,
    author = "Bazak, Sylwia and Mrowczynski, Stanislaw",
    title = "{Production of $^4\mathrm{Li}$ and $p\!-\!^3\mathrm{He}$ correlation function in relativistic heavy-ion collisions}",
    eprint = "2001.11351",
    archivePrefix = "arXiv",
    primaryClass = "nucl-th",
    doi = "10.1140/epja/s10050-020-00198-6",
    journal = "Eur. Phys. J. A",
    volume = "56",
    number = "7",
    pages = "193",
    year = "2020"
}

\end{document}